# Tunable Frequency Comb Generation from a Microring with a Thermal Heater

**Xiaoxiao Xue[1*], Yi Xuan[1,2], Pei-Hsun Wang[1], Jian Wang[1,2], Dan E. Leaird[1], Minghao Qi[1,2], and Andrew M. Weiner[1,2]**

[1]*School of Electrical and Computer Engineering, Purdue University, 465 Northwestern Avenue, West Lafayette, Indiana 47907-2035, USA*

[2]*Birck Nanotechnology Center, Purdue University, 1205 West State Street, West Lafayette, Indiana 47907, USA*

**xue40@purdue.edu*

**Abstract:** We demonstrate a novel comb tuning method for microresonator-based Kerr comb generators. Continuously tunable, low-noise, and coherent comb generation is achieved in a CMOS-compatible silicon nitride microring resonator.

Optical frequency comb generation based on Kerr effect in nonlinear microresonators is very attractive for its simplicity, small size, and potential of chip-level integration [1-2]. For many applications, the tunability of the generated combs is required to perform some essential functions, such as aligning the comb lines with the channels in WDM communication systems, or achieving stabilization guided by self-referencing with a *f*-2*f* interferometer. One method of tuning, namely simply adjusting the pump frequency and utilizing the thermal self-locking effect, has been demonstrated for a fused silica microresonator [3]. But generally combs are not expected to maintain a uniform low-noise state in the tuning process, because the intensity noise and coherence of Kerr combs sensitively depend on the phase detuning between the resonance and the pump frequency [4]. Here we demonstrate a novel comb tuning scheme incorporating a thermal heater. Continuously tunable, low-noise and coherent comb generation is achieved in a CMOS-compatible silicon nitride (SiN) microring. The experimental setup is shown in Fig. 1. A continuous-wave light is injected into a SiN microring for comb generation. The comb is then line-by-line shaped by a pulse shaper to form a bandwidth-limited pulse train [5]. A fraction of the comb power is detected by a photodetector to monitor the intensity noise. The microscope image of the microring is shown in the inset of Fig. 1. The ring radius is 100 μm. The waveguide cross-section dimension is $2\ \mu m \times 550\ nm$ which corresponds to a normal dispersion regime. A heater made of Au is fabricated on top of the ring for thermal tuning. A 3.5-μm $SiO_2$ layer is deposited between the SiN layer and the metal layer to avoid absorption loss.

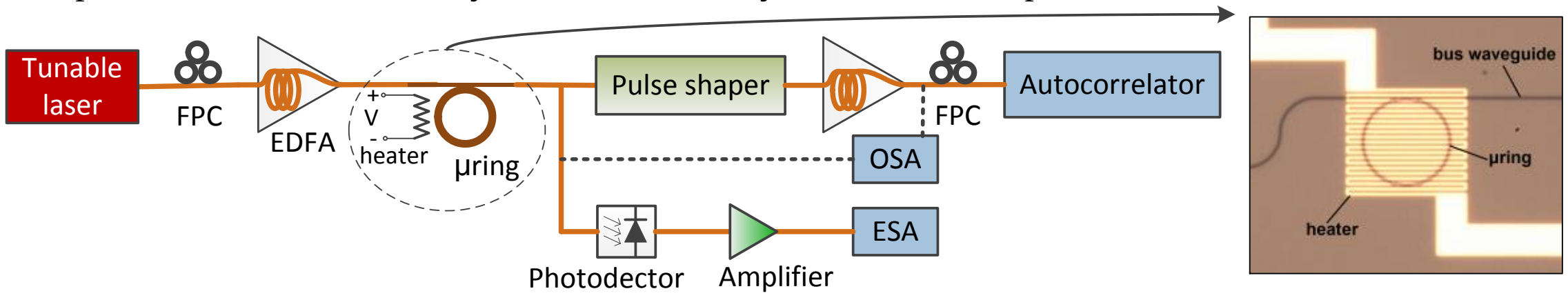

Fig. 1. Experimental setup of the tunable Kerr comb generation. FPC: fiber polarization controller; EDFA: Erbium-doped fiber amplifier; μring: silicon nitride microring; OSA: optical spectrum analyzer; ESA: electrical spectrum analyzer;

The microring resonance can be red-shifted by increasing the voltage applied to the heater because of the thermal-optic effect. The normalized transmission of the microring in the 1545-1555 nm range is shown in Fig. 2(a). There are two mode families, and the one with higher quality factor ($Q_{Loaded} = 7\times10^5$) is used for comb generation. The free spectral range of this mode is 1.85 nm. Figure 2(b) shows one resonance shifted by varying the heater voltage; and figure 2(c) shows the resonance shift versus the heater power. The heater resistance is 291 ohm. The resonance can be shifted by one FSR when the heater power is 2.2 W (corresponding to a voltage of 25.5 V).

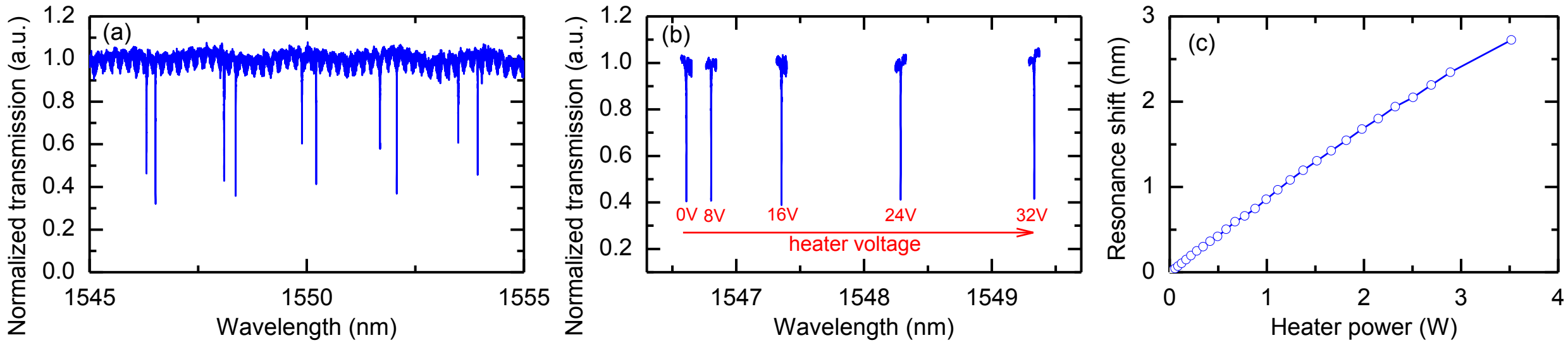

Fig. 2. (a) Normalized transmission of the microring; (b) one resonance shifted by heater control; (c) resonance shift versus heater power.

For comb generation there are now two ways to push the pump frequency into the resonance. The first way is the conventional one by tuning the pump from higher frequency to lower frequency. The second way which is achievable with the thermal tuning scheme is by tuning the resonance. In this case, the resonance is first thermally tuned to the red side and then slowly tuned back. Because of the thermal-locking effect and Kerr effect, the resonator will be locked to the pump frequency when it approaches the pump in the back tuning process. This novel method makes it possible to generate combs with a fixed-wavelength pump laser which is simpler than a finely tunable laser. The inset 1 of Fig. 3 shows the comb spectrum generated by using this resonance tuning method when the pump is fixed at 1549.3 nm. The resonance centered at 1548.365 nm is used for comb generation. The heater voltage is first increased from 0 V to 20 V and then slowly reduced backwards. The comb transitions to a low-noise state when the pump power is optimized to 1.31 W and the heater voltage 8.57 V. By changing the heater voltage and the pump frequency in tandem, the spectrum of the low-noise comb can be continuously tuned. Figure 3 shows the tunable comb spectra when the pump wavelength is tuned from 1549 nm to 1550.4 nm with a step of 0.1 nm. The comb keeps a nearly identical spectral shape in the tuning process. The demonstrated tuning range is 1.4 nm ( $0.76\times \mathrm{FSR}$ ). We believe that a tuning range more than one FSR can be easily achieved by using the thermal-tuning scheme with improved heater designs. To probe the coherence between the comb lines, a fraction of the comb spectrum is selected and shaped line-by-line both in amplitude and phase by a pulse shaper to form a bandwidth-limited Gaussian pulse train [5]. The shaped comb spectrum when the pump is 1549.3 nm is shown in the inset 2 of Fig. 3. Figure 4 shows the measured 15 autocorrelation traces of the tunable combs after line-by-line shaping; each corresponds to one comb with a different pump wavelength. All the autocorrelation traces agree quite well with the theoretical result which is calculated by assuming perfect phase compensation. The autocorrelation width is 318 fs FWHM, which implies an intensity width of ~200 fs FWHM. Figure 5 shows the comb intensity noise spectra which approach the sensitivity of the electrical spectrum analyzer. Both the good autocorrelation results and the low intensity noise suggest high coherence between the comb lines

In conclusion, we have demonstrated a novel method for tunable Kerr comb generation. Continuously tunable, low-noise, and coherent comb generation is achieved in a CMOS-compatible SiN microring with a thermal heater.

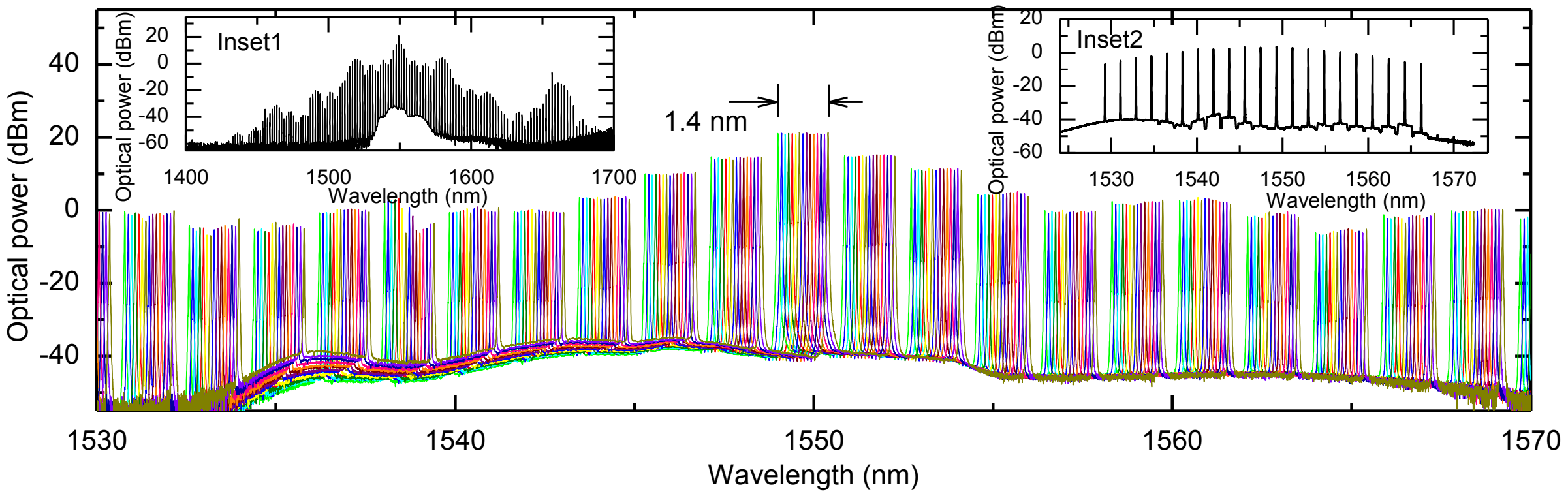

Fig. 3. Tunable comb spectra. Inset.1 & inset.2: full-range spectrum & shaped spectrum when the pump is 1549.3 nm.

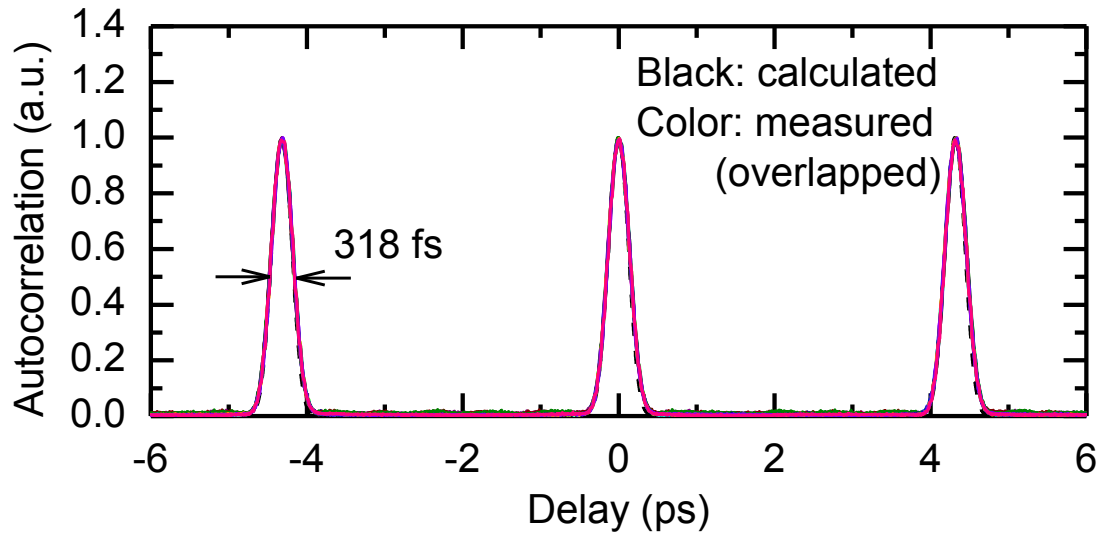

Fig. 4. Autocorrelation traces of the tunable combs.

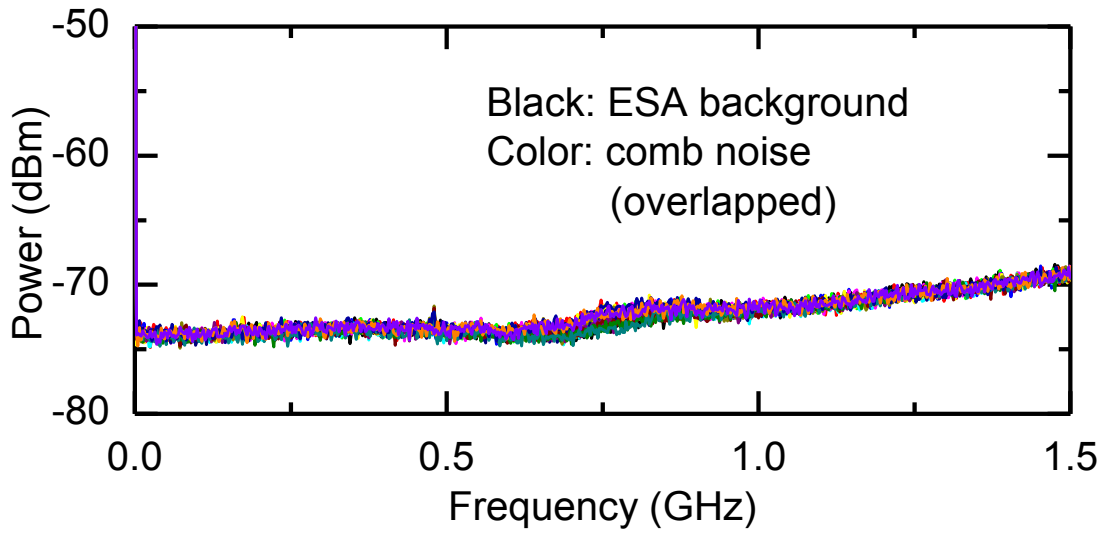

Fig. 5. Intensity noise spectra of the tunable combs.

## References

[1] T. J. Kippenberg, R. Holzwarth, and S. A. Diddams, Science, **332,** 555-559 (2011).

[2] J. S. Levy, A. Gondarenko, M. A. Foster, A. C. Turner-Foster, A. L. Gaeta, and M. Lipson, Nat. Photonics **4,** 37-40 (2010).

[3] P. Del'Haye, T. Herr, E. Gavartin, M. L. Gorodetsky, R. Holzwarth, and T. J. Kippenberg, Phys. Rev. Lett. **107,** 063901 (2011).

[4] T. Herr, K. Hartinger, J. Riemensberger, C. Y. Wang, E. Gavartin, R. Holzwarth, M. L. Gorodetsky and T. J. Kippenberg, Nat. Photonics, **6,** 480-487 (2012).

[5] F. Ferdous, H. Miao, D. E. Leaird, K. Srinivasan, J. Wang, L. Chen, L. T. Varghese, and A. M. Weiner, Nat. Photonics **5,** 770-776 (2011).